\documentclass{desyproc}

\begin{document}
\title{The Future of Low Energy Photon Experiments}

\author{{\slshape Axel Lindner$^1$}\\[1ex]
$^1$DESY, Notkestra{\ss}e 85, 22607 Hamburg, Germany}

\contribID{17}

\confID{1407}  
\desyproc{DESY-PROC-2009-03}
\acronym{PHOTON09} 
\doi  

\maketitle

\begin{abstract}
``Light-shining-through-a-wall'' experiments search for Weakly Interacting Sub-eV Particles (WISPs). The necessity and status of such enterprises as well as their future potential are sketched. 
\end{abstract}

\section{An Experimentalist's Motivation}

Evidence is mounting that the known constituents of matter and forces beautifully summarized in the Standard Model do not fully describe the world around us. Such arguments arise from astrophysical and cosmological observations as well as from theoretical considerations. There are strong convictions among scientists that new experiments at the high energy frontier at LHC will provide insight into physics beyond the Standard Model. Although theoretically well motivated, focusing the search for new physics onto highest available energies neglects evidences pointing at the opposite energy scale. Extensions of the Standard Model may manifest itself also at meV energy scales, nine orders of magnitude below the mass of the electron (see contribution of J.~J\"ackel to these proceedings).

Generally, new very light and very weakly interacting particles denoted as WISPs (Weakly Interacting sub-eV Particles) occur naturally in string theory motivated extension of the Standard Model~\cite{Goodsell:2009xc,Ringwald:2008cu,Ahlers:2007rd}. 
There could be bosons and fermions, charged and uncharged particles~\cite{Ringwald:2008cu}.
WISPs may interact with ordinary matter via the exchange of very heavy particles related to very high energy scales and thus give insight into physics at highest energy scales. The reader is referred to~\cite{Ehret:2009sq} and references therein for a more detailed view.
One prime example for a WISP is the QCD axion \cite{Weinberg:1977ma,Wilczek:1977pj} invented to explain the CP conservation of the strong interaction. From astrophysical observations its mass should be below about 1~eV. For the QCD axion such a low mass implies very weak interactions with the other constituents of the Standard Model (see~\cite{Amsler:2008zzb}).
It is striking that a QCD axion with a mass around 1~$\rm \mu$eV is a perfect candidate for cold dark matter in the Universe~\cite{Abbott:1982af,Sikivie:2009fv}. 
A discovery of the axion could solve long lasting questions of particle physics and cosmology simultaneously. It is worthwhile to note that also dark energy might be attributed to new physics at the meV scale~\cite{Masso:2009ms}.

Interestingly, very different astrophysical observations suggest the existence of very light axion-like particles. The cooling of white dwarfs can be modeled significantly better, if an additional energy loss due to axion-like particles is taken into account~\cite{Isern:2008fs}.
Luminosity relations of active galactic nuclei (AGN) show patterns which are best understood if axion-like particles do exist~\cite{Burrage:2009mj}. 
The surprisingly high transparency of the Universe to TeV photons from AGNs at cosmological distances may be explained by back and forth oscillations of photons into axion-like particles~\cite{Roncadelli:2009zz,SanchezConde:2009wu}.
Similarly, such oscillations might explain the candidate neutral ultra-high-energy particles from distant BL Lac type objects ~\cite{Fairbairn:2009zi}.
The heating of the solar corona is not understood, but may be attributed to an energy flow mediated by axion-like particles~\cite{Zioutas:2009bw}.

\section{WISP Searches}
At present for most of the WISP species the most stringent limits on their existence originate from astrophysics considerations. In general, the existence of WISPs would open up new energy loss channels for hot environments in stars for example and thus shorten lifetimes or cooling cycles~\cite{Amsler:2008zzb,Gninenko:2008pz}. They could also show up in analyses of the cosmic microwave background radiation~\cite{Jaeckel:2008fi,Burrage:2009yz,Mirizzi:2009iz,Mirizzi:2009nq}.

In addition direct searches for axion-like particles produced in the sun~\cite{Zioutas:2009bw} or as constituents of galactic dark matter~\cite{Duffy:2006aa} have greatly progressed in recent years and reached impressive sensitivities. 

However, interpretations of astrophysics data are always hampered by the uncontrolled production mechanism of WISPs. Effective theories have been presented, where the production of some WISP species is suppressed in hot environments~\cite{Jaeckel:2006xm,Redondo:2008en}. If such scenarios are true, astrophysics experiments might fail to detect WISPs while laboratory experiments could open up this new physics window.
Literally, astrophysics deals with ``astronomical'' or ``cosmological'' distances on the one hand and microscopic distances in hot dense plasmas on the other hand for example. Intermediate distances are only probed in the laboratory.

The are numerous experimental efforts to probe for WISPs in the laboratory. Typically they are searched for by looking for new effects in gravitational or QED environments. The latter one comprises atom physics like Lamb shift, positronium decay, Casmir forces or photon-photon interaction.

\section{Photon-photon Interactions and \\ ``Light-Shining-through-a-Wall''}
\begin{figure}[tbh]
\centerline{\includegraphics[width=0.17\textwidth,angle=-90]{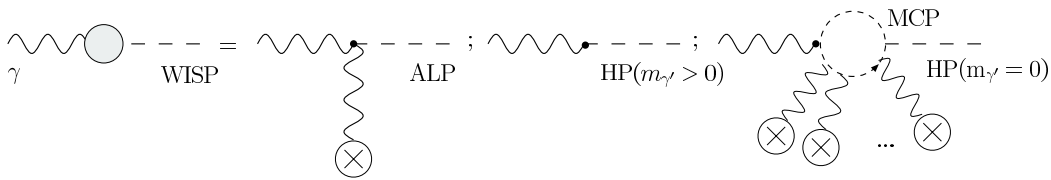}}
\caption{A collection of some Feynman diagrams responsible for the mixing term between photons and different hypothetical ``weakly interacting sub-eV particles'' (WISPs). 
Photon oscillations into Axion-like particles (ALPs) and massless hidden photons (HPs) via mini-charged particles (MCP) require the presence of a background electromagnetic field, denoted by crossed circles.} 
\label{fig:blobs}
\end{figure}
Since the discovery of the positron there is profound interest in light-light interaction, because two photons can couple via virtual electron-positron pairs. 
If an intense laser beam is shone into a strong magnetic field, the interaction of the laser light with the (virtual) photons of the magnetic field gives rise to the so-called magnetic vacuum birefringence~\cite{Heisenberg:1935qt}: light polarized perpendicular to the magnetic field moves a little slower than photons polarized perpendicular.
However, the effect is very tiny ($\rm n_\perp - n_\parallel  =  3.6\cdot10^{-22}$ for B~=~9.5~T) and has escaped experimental verification since decades. Present day experiments have to be improved by about three orders of magnitude to verify this QED prediction.
A similar phenomenon is the so-called photon splitting. Here the original photon vanishes with a probability depending on its polarization with respect to the magnetic field orientation. 
This reminds of a crystal's dichroism.
Unfortunately, QED predicts only a very weak effect most likely beyond experimental reach for the foreseeable future (see also contribution of G.~Cantatore to these proceedings).

The reason for the weakness of both effects is the high mass of the electron as compared to the eV photons.
If WISPs with much smaller masses exist and couple to photons, they could give rise to the above mentioned two effects with amplitudes well above the QED expectations~\cite{Ahlers:2008jt,Cantatore:2008ju}. 
Fig.~\ref{fig:blobs} (see~\cite{Ehret:2009sq}) displays Feynman diagrams for the production of some WISP flavors due to the interaction of photons.
WISPs could also be detected in a very convincing 
\begin{wrapfigure}{l}{0.5\textwidth}
\centerline{\includegraphics[width=0.5\textwidth,angle=0]{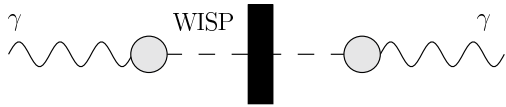}}
\caption{Sketch of a ``light shining through a wall experiment''. The gray blob indicates the mixing term between photons and the ``weakly interacting sub-eV particles'' (WISPs).} 
\label{fig:lsw}
\end{wrapfigure}
and spectacular manner with the so called by ``light-shining-through-a-wall'' (Fig.~\ref{fig:lsw}, see~\cite{Ehret:2009sq}). 
In the first part of an experiment WISPs are produced from intense laser light, either by interaction with a strong magnetic field or by kinetic mixing. This first part is sealed-in by a light-tight wall from the second part. Only WISPs can traverse the wall due to their very low cross-sections. Behind the wall they could convert back into photons with exactly the same properties as the light generating the WISPs. This gives the impression of ``light-shining-through-a-wall'' (LSW).
This article will focus on the status and future of LSW experiments. Such installations offer the possibility to significantly increase sensitivities in future.

\section{Present and Future LSW Experiments}
In the following production and detection of axion-like particles (ALPs) will be considered in some detail. 
ALPs are of prime interest due to their similarity with the QCD axion, but please keep in mind that many other species (see above) may populate the WISP-zoo, which further enhances the discovery potential of LSW experiments.

\subsection{Axion-Like Particles}  
Axion-like particles, ALPs, denote here neutral scalar or pseudoscalar particles coupling to electromagnetic fields similar to the QCD axion.  
If light is shone into a magnetic dipole field scalar (pseudoscalar) ALPs can be produced if the light is polarized perpendicular (parallel) to the direction of a magnetic dipole field.
Assuming a symmetric set-up of an LSW experiment with the magnetic field length BL before and after the wall the photon-ALP-photon conversion probability reads 
\begin{equation}
\label{lswprob}
P(\gamma \to \phi \to \gamma) = \frac{1}{16 \beta_\phi^2}(gBL)^4 \left( \frac{\sin\frac{1}{2}qL}{\frac{1}{2}qL} \right)^4,
\end{equation}
where $\beta_\phi$ denotes the velocity of the ALP and $q = p_\gamma - p_\phi$.  
The probability rises with $(BL)^4$ resulting from the coherent ALP-photon-ALP conversions.
For a more detailed discussion of these effects the reader is referred to~\cite{Ehret:2009sq} and the references therein.

Note that for $qL<<1$ Eq.\ref{lswprob} reduces to
\begin{equation}
\label{lswprobred}
P(\gamma \to \phi \to \gamma) = \frac{1}{16 \beta_\phi^2}(gBL)^4.
\end{equation}
If $qL<<1$ does not hold anymore the conversion probability drops and the limits on the coupling g worsens as visible in Fig.~\ref{fig:lswreach}.
This effect can be made lively by imaging an overlay of the wave functions of photons and ALPs. If the ALP's momentum decreases, its wavelength rises and hence runs out of phase compared to the photon wave function.
The first LSW experiment was carried through by the BFRT collaboration~\cite{Cameron:1993mr}. Table \ref{tab:lswexp} summarizes present day experiments. 
\begin{table}[htb]
\setlength{\extrarowheight}{4pt}
\centerline{\begin{tabular}{|l|l|r|m{2.5cm}|}
\hline
Experiment            & \ \ \ \ \ Number of photons & Magnet $B\cdot L$ &  g [$10^{-7} \rm{GeV^{-1}}$],\newline 95\% CL limit \\\hline
ALPS @ DESY~\cite{Ehret:2009sq}    & $5\cdot10^{24}$, 532~nm, cw         & 22+22~Tm    &  \ \ \ \ \ \ \ \ $4.1$ \\\hline
BMV @ Toulouse~\cite{Fouche:2008jk} & $6\cdot10^{23}$, 1060~nm, 82 pulses & 4.4+4.4~Tm  &  \ \ \ \ \ \ \ \ $10$ \\\hline
GammeV @ FNAL~\cite{Chou:2007zzc}  & $6\cdot10^{23}$, 532~nm, 5~Hz pulses& 15+15~Tm    &  \ \ \ \ \ \ \ \ $2.9$ \\\hline
LIPSS @ JLAB~\cite{{Afanasev:2008jt}}   & $6\cdot10^{25}$, 935~nm, FEL        & 1.8+1.8~Tm  &  \ \ \ \ \ \ \ \ $12$ \\\hline
OSQAR @ CERN~\cite{osqarspsc:2007}   & 488+514~nm, cw     & 136+136~Tm  & \ \ \ \ \ \ \ \ $3.4$ \\\hline
\hline
\end{tabular}}
\caption{Table of present day LSW experiments. The second column lists the total number of photons used for the analyses, the last column shows typical sensitivities achieved. More details are given in the references shown.}
\label{tab:lswexp}
\end{table}

For a typical experimental set-up similar to the ALPS experiment at DESY or GammeV at FNAL Eq.~\ref{lswprobred} reads
\begin{equation}
\label{lswprobexp}
P(\gamma \to \phi \to \gamma) = 9.60\cdot10^{-25} \left( \frac{g}{10^{-7}\rm{GeV^{-1}}} \frac{B}{5\rm{T}} \frac{L}{4\rm{m}} \right)^4.
\end{equation}
It is evident that a very large flux of photons is required to probe for ALPs in a LSW experiment. 
Table \ref{tab:lswpar} summarizes the main experimental parameters, the dependence of the sensitivity in g on these parameters and possible future improvements. As a benchmark the ALPS experiment at DESY in Hamburg is used. 
\begin{table}[hbt]
\setlength{\extrarowheight}{4pt}
\centerline{\begin{tabular}{|l|r|c|c|}
\hline
Parameter & g dependence & ALPS~aims & future exp.  \\\hline
Laser power (cw) & $g\propto P^{-\frac{1}{4}}$ & $P$=1kW & $P$=100kW \\\hline
Magnetic field & $g\propto (BL)^{-1}$ & $BL$=22Tm  & $BL$=600Tm \\\hline
Detector sensitivity  & $g\propto \epsilon^{\frac{1}{4}}$ & $\epsilon=4$mHz & $\epsilon$=0.02mHz \\\hline
Measurement time & $g\propto t^{-\frac{1}{8}}$ & $t$=25h & $t$=25h \\\hline
\hline
\end{tabular}}
\caption{Table of the main parameters of a LSW experiment searching for ALPs. The second column gives the sensitivity of limits on the photon-ALP coupling on the experimental parameters. The third column lists the corresponding parameters aimed for at the ALPS experiment at DESY, while the forth column shows parameters of a possible future experiment.}
\label{tab:lswpar}
\end{table}

In the near term laboratory LSW experiments aim for reaching a sensitivity for the photon-ALP coupling of $g = 10^{-7}\rm{GeV^{-1}}$.
It's worth mentioning that this probes for new physics (if WISP exist) at the 100 TeV scale\footnote{The axion to photon coupling g is given by $g= \alpha \cdot g_\gamma/\pi f_\alpha$, where $f_\alpha$ denotes the new energy scale and $g_\gamma$ a factor derived from theory expected to vary by about an order of magnitude for the QCD axion.} already.   
However, at present astrophysics limits on the coupling of ALPs to photons are about 3 orders of magnitude more restrictive.
On the long run laboratory experiments should strive for surpassing these limits.

\subsubsection{Laser beam power}  
Usually LSW experiments shine laser light through long and tight magnet bores. At ALPS the open aperture amounts to only 16~mm~\cite{Ehret:2009sq} for a total length of about 17~m.
Therefore the demands on the beam quality are rather high.
Given these constraints optical or infrared lasers with cw powers of a few 10 W at maximum are available resulting in a photon number flux of roughly $\rm 10^{20}$~Hz.
This can be improved significantly by recycling the photons with the help of an optical delay line (used by BFRT~\cite{Cameron:1993mr}) or by setting up a resonant optical cavity.
This has been realized for the first time by the ALPS experiment~\cite{Ehret:2009sq}. Improvements of the ALPS resonator are under way aiming for an effective power of 1~kW from 4~W of primary laser power.
Considering the laser R\&D for gravitational wave antennas~\cite{Luck:2006ug,The07} the next decade could result in primary lasers with powers of the order of 100~W.
This power might be enhanced in an optical cavity by 3 orders of magnitude resulting in an effective power around 100~kW.
Hence the mid term future might give effective photon fluxes close to $\rm 10^{24}$~Hz.
Further significant improvements will demand focused R\&D programs.

Some experiments use or have used different set-ups with pulsed lasers.
This is a very reasonable approach as long as the photon detector sensitivity can be enhanced by triggering to suppress background noise (GammeV~\cite{Chou:2007zzc}) or if very strong pulsed magnets are used (BMV~\cite{Fouche:2008jk}).
This is not discussed further in this article.

\subsubsection{Magnets}  
ALPS is using a spare HERA dipole magnet. 
Half of the dipole is used for generating WISPs, the second half for converting them back into photons.
This gives $BL = 23$~Tm for both processes. 
OSQAR at CERN is setting up an installation with one LHC dipole each for the generation and regeneration part resulting in 139~Tm.
Four LHC dipoles or two DLHC dipole each would provide about 600~Tm. Such a set-up might be realized with limited costs in the mid-term future.
Going beyond would require a considerable amount of resources for magnet developments and/or constructing very long dedicated cryogenic magnet stands for WISP experiments.
Clearly such an enterprise would be worthwhile, because the sensitivity in $g$ improves linearly with the magnetic field strength and length (see Table \ref{tab:lswpar}).
For the time being $BL = 600$~Tm is assumed for the future, a factor of 26 compared to the present-day ALPS.

\subsubsection{Detector sensitivity}
Most of the present day LSW experiments use commercial CCD cameras to search for reconverted photons from WISPs behind the wall. 
Because the light looked for has the same properties as the laser light shown into the experiment one can profit from the usually outstanding beam quality by focusing the beam spot onto very few pixels of the CCD.
Signal areas smaller than 50x50~$\rm \mu m^2$ are regularly achieved.
An optical resonator in front of the wall does not harm, but improves this approach (while this strong focusing could hardly be realized with an optical delay line due to geometrical reasons).
With a typical dark current of $\rm 10^{-3}$ electrons per pixel and second, a read-out noise of 3.8 electrons and a signal region of 3x3 pixels one expects for one hour exposures a RMS of only 12 electrons. 
Taking into account an overall photon detection efficiency of roughly 60\% this translates into a RMS of 20 photons (assuming a gain factor of 1 electron/photon).
Hence one expects a 95\%CL photon flux limit around 4~mHz for about 10 one hour exposures (the typical order of magnitude for measurement times in LSW experiments).
This corresponds to an energy flux of $\rm 2\cdot 10^{-21}$~W for 532~nm photons.

In future this sensitivity might be enhanced considerably by using transition edge sensors (TES)~\cite{TES1,Romani:2002qn}. 
Here a sensor is cooled down to about 100~mK and operated in the transition region between a superconducting and normal conducting state. 
Due to the very low heat capacity of such a state the energy deposit of a single photon results in a significant temperature rise and is well measurable. 
TES allow for essentially background free counting of individual photons.
Only radioactivity or cosmic ray interactions would provide some remaining background. 
From studies with many dark frames taken with the ALPS CCD the background rate is estimated as roughly 0.02~mHz in a 50x50~$\rm \mu m^2$ signal.
Hence it should be possible in future to set-up a detector system sensitive to an energy flux of at least two orders of magnitude below the above mentioned number, i.e. $\rm 10^{-23}$~W for 532~nm photons.
This might be further improved by shielding and proper selection of materials.
In addition TES detectors register the arrival time of individual photons and coarsely estimate their energies.   
This is a clear advantage compared to present day long exposure CCD frames.
\\ 
\\ 
The improvements listed above are judged as likely possibilities to be reached within the next few years. 
According to Table~\ref{tab:lswpar} $ g = 10^{-7}\rm{GeV^{-1}}$ as aimed for at ALPS is expected to be multiplied by a factor of
$\rm \left({100 kW}/{1 kW}\right)^{-\frac{1}{4}} \cdot \left({600 Tm}/{23 Tm}\right)^{-1} \cdot \left({0.02 mHz}/{4 mHz}\right)^\frac{1}{4} = 0.003.$
Couplings significantly below $\rm 10^{-9}GeV^{-1}$ would be accessible in LSW experiments.
However, a gap of about an order of magnitude would still to be bridged to surpass present-day limits from astrophysics. A second optical resonator is a promising means to tackle this challenge.
 
\subsubsection{Regeneration cavity}
The idea of a resonantly enhanced axion photon regeneration was put forward first in 1993 by F.~Hoogeveen and T.~Ziegenhagen~\cite{Hoogeveen:1990vq} and independently rediscovered in 2007 by P.~Sikivie, D.B.~Tanner and K.~ van~Bibber~\cite{Sikivie98}.
The basic idea is to set-up an optical resonator also in the regeneration part of a LSW experiment very similar to the optical resonator in the first part.
The second resonator effectively increases the conversion probability of a WISP into a photon. 
To understand this consider that the freely propagating WISP related wave behind the wall of the LSW experiment comprises a very tiny electromagnetic photon component. 
Due to this small component the WISP might convert into a real photon.
An optical resonator enhances this small component in the same way as the wave amplitude for real photons is increased.
The transition probability of WISP to photons rises with the power amplification factor of an optical resonator in the second part of the LSW experiment behind the wall.
Consequently the sensitivity of such a set-up for the coupling constant $g$ improves with the square root of this factor.
The technical challenge is to lock the second cavity to exactly the same frequency and the same mode as the first cavity (used to enhance the effective laser photon 
\begin{wrapfigure}{l}{0.5\textwidth}
\begin{center}
\vspace*{-0.8cm}
 \includegraphics[width=0.5\textwidth,angle=0]{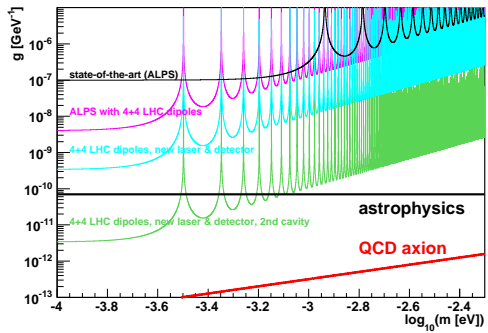}
\caption{The sensitivity aimed for at present day LSW experiments on the coupling of ALPs to photons vs. the ALPs mass as well as future experiments with stronger magnets, improved laser and detector system and an additional regeneration cavity (from top to bottom). The horizontal line indicates today's limits derived from astrophysics. Always the regions above the lines are excluded. The slanted line shows the order of magnitude expected for the coupling strength of the QCD-axion to photons.} 
\vspace*{-0.5cm}
 \label{fig:lswreach}
\end{center}
\end{wrapfigure}
flux).
Obviously one can not use laser light of the same wavelength to produce WISPs and to lock the regeneration cavity without spoiling the sensitivity to detect photons from reconverted WISPs.
One possibility rests on a two colored laser. 
Such an installation is being used in the ALPS experiment at DESY, where part of the 1064~light from an infrared laser is frequency doubled to 532~nm~\cite{Ehret:2009sq}.
In future the 532~nm light could be used to lock the generation and regeneration cavity while the (more powerful) 1064~nm radiation is used to produce WISPs and to search for regenerated photons behind the wall.
A detector can be shielded from the 532~nm radiation leaving the regeneration cavity by properly suited bandpass filters (narrow bandwidth filters with 75\% transmission at 1064~nm and optical densities larger than 5 at other wavelenths are easily available).
Efforts to realize such a set-ups are starting in the US~\cite{Mueller:2009wt} and Europe.  
With a regeneration cavity it should be possible to improve the sensitivity of LSW experiments by another two orders of magnitude to reach $g = 10^{-11}\rm{GeV^{-1}}$ and lower values.
Fig.~\ref{fig:lswreach} displays the reach of LSW experiments exploiting the power of a regeneration cavity. Present day limits derived from astrophysics are likely to be surpassed in future.
However, probing for the QCD-axion in the laboratory will probably remain a challenging target for quite some time.  

\section{Further Ideas}
\begin{wrapfigure}{r}{0.5\textwidth}
\begin{center}
\vspace*{-0.3cm}
 \includegraphics[width=0.4\textwidth,angle=0]{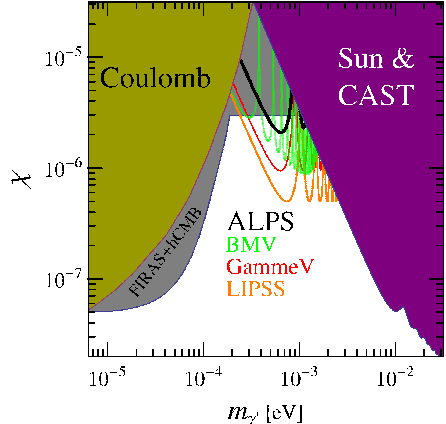}
\vspace*{-0.5cm}
\caption{Present day limits for the mixing of photons with their massive counterparts from a ``hidden sector''. Laboratory experiments ~\cite{Ehret:2009sq,Fouche:2008jk,Chou:2007zzc,Afanasev:2008jt,Ahlers:2007qf} surpass astrophysics limits the mass region around 1~meV already now.}
\end{center}
\vspace*{-1.8cm}
\label{fig:hp}
\end{wrapfigure}

For illustration purpose the argumentation above has concentrated on axion-like particles.
It should be stressed again that these are only one kind in a hypothetical zoo of WISPs. 
Another very interesting species are massive photons (dubbed ``HP'', because they might originate from the ``hidden sectors'' of string theory for example) with very weak couplings to Standard Model particles.
If HPs exist they oscillate with photons back and forth in vacuum very similar to neutrino flavor mixing.
Hence one can search for HPs also in LSW experiments, but without the necessity of an external electromagnetic field.
The present experimental limits are shown in Fig.~\ref{fig:hp}. It is interesting to note that the mass scale around 1~meV is hardly probed by astrophysics.
Hence any improvement in laboratory based HP searches probes unknown territory in the
parameter space.
The improvements of lasers and detectors as well as the usage of a regeneration cavity could improve limits on the mixing parameter by more than two orders of magnitude.

The potential of LSW experiments at synchrotron sources~\cite{Dias:2009ph}
and with microwave cavities~\cite{Caspers:2009cj}, the new possibilities at free electron lasers~\cite{Ringwald:2001cp,Rabadan:2005dm}
or ``current-through-a-wall'' experiments~\cite{Gies:2006hv} could not be addressed here. 
The reader is referred to the references given.

\section{Summary}
We witness a revival of experiments at the low energy frontier searching for new, very light and very weakly interacting particles. This complements enterprises at the high energy and intensity frontiers.
In the recent years various investigations have shown that the low energy frontier indeed may offer a window to physics beyond the Standard Model. This is strongly supported by theory, giving rise to a zoo of WISPs in addition to the QCD-axion, and interesting observations from astrophysics.  

Laboratory experiments have the potential to surpass the reach of astrophysics in the search for WISPs. 
Some concentrated R\&D efforts on laser technology, long optical resonators and single photon detectors are required to meet this aim. 
These can be achieved by new collaborations among the high energy physics and laser interferometer communities (operating gravitational wave antennas) for example.  
Firm bases exist and no major show-stoppers are envisaged today.  

Present day and future LSW experiments probe nature in a regime of very weak couplings and very low masses. 
In addition efforts should be strengthened to test the validity of the Standard Model in very strong fields (to test non-linear QED for example).
The field of particle physics as a whole can only gain if such small scale experiments complement the physics searched for in billion dollar installations at the high energy or intensity frontiers.

\section{Acknowledgements}
Many thanks to the organizers of PHOTON 2009 for giving me the opportunity to contribute to this interesting meeting with a broad physics scope! I thank my colleagues of the ALPS collaboration as well a G.~Cantatore and J.~J\"ackel for stimulating and fruitful discussions.

 \begin{footnotesize}

\end{footnotesize}

\end{document}